%% file: lista_hepex.tex
\documentclass[12pt]{article}
\usepackage{epsfig}

\input econfmacros.tex

\def\Title#1{\begin{center} {\Large {\bf #1} } \end{center}}

\begin{document}

\def\stat{({\mathrm s}{\mathrm t}{\mathrm a}{\mathrm t}.)}
\def\syst{({\mathrm s}{\mathrm y}{\mathrm s}{\mathrm t}.)}
\def\theor{({\mathrm t}{\mathrm h}.)}
\def\GeV{{\mathrm G}{\mathrm e}{\mathrm V}}
\def\MeV{{\mathrm M}{\mathrm e}{\mathrm V}}

\begin{flushright}
BABAR-PROC-04/133\\
SLAC-PUB-10875\\
\end{flushright}

\Title{Review of Recent BaBar Results}

\bigskip\bigskip


\begin{raggedright}  

{Luca Lista\index{Lista, L.}\\
{\footnotesize representing the BaBar Collaboration}\\
INFN Sezione di Napoli\\
Complesso Universitario di Monte Sant'Angelo, via Cintia\\
I-80126 Napoli, ITALY}
\bigskip\bigskip
\end{raggedright}

{\small Proceedings of: 2$^{{\mathrm n}{\mathrm d}}$ International Conference in High-Energy Physics: HEP MAD 04,
26 Sep - 4 Oct 2004, Antananarivo, Madagascar}

\section{Abstract}

We present a review of recent results from BaBar experiment. BaBar
detector has collected about 256 millions of $B\bar B$ events at PEP-II, 
the asymmetric $e^+e^-$ collider located at SLAC 
running at the $\Upsilon(4S)$ resonance. 
We have studied CP violation in
B mesons, observing the first evidence of direct CP violation in B meson
decays and measured CP asymmetries relevant for the determination of the angles of the CKM 
Unitarity Triangle. 
BaBar physics program covers many other topics, including measurements of
CKM matrix elements, charm physics, and search for new physics processes.

\section{Introduction}

BaBar experiment runs at the PEP-II asymmetric B-factory located at SLAC 
laboratories. The center of mass energy corresponds to the mass of the $\Upsilon(4S)$ resonance
which decays predominantly into pairs of $B$ and anti-$B$ mesons. The experiment has
recorded until July 31$^{{\mathrm s}{\mathrm t}}$ 2004 an integrated luminosity of about 244~fb$^{-1}$, with a 
peak luminosity of $9.2\times 10^{33}$~cm$^{-2}$s$^{-1}$. 

A detailed description of the detector can be found elsewhere~\cite{BaBar}.
A silicon vertex tracker (SVT) consisting of five layers and a
drift chamber (DCH) with 40 stereo layers
provide the detection of charged particles
whose momentum is measured in a 1.5-T solenoidal magnetic field.
The energy loss measurement (${\mathrm d}E/{\mathrm d}x$) in the 
tracking detectors contributes to the charged particle identification.
A detector of internally reflected Cherenkov radiation (DIRC) 
is the main subsystem used for charged hadron identification.
A finely segmented CsI(Tl) electromagnetic calorimeter (EMC) 
is used to measure electron and photon energy.
The instrumented flux return (IFR) segmented with
Resistive Place Chambers provides muon identification.
Neutral hadrons ($K_L$) are identified in the EMC and the IFR.

\section{CP Violation}

Violation of the CP symmetry can be explained in the Standard Model
with a non eliminable phase in the Cabibbo-Cobayashi-Maskawa quark mixing matrix.
It has been first observed in the neutral Kaon system as the effect
of CP violation in mixing. This type of CP violation is expected to be small 
($\sim 10^{-3}\div 10^{-4}$) in the neutral $B$-meson system. A large violation
is possible in the Standard Model both as direct CP violation and 
as time-dependent CP violation in the interference between mixing and decay.

\subsection{Direct CP violation}

A significant direct CP violation may arise in the decay
$B^0 \rightarrow K^-\pi^+$ from the interference
between the tree and penguin diagrams. We have reconstructed 
$B^0$ decaying  to $K^-\pi^+$ and its charge conjugate
mode; $B$ meson decays are identified using the kinematical variables:
\begin{eqnarray}
\Delta{E} & = & E_B^{*} - \sqrt{s}/2 \,\, ,\\
m_{ES} & = & \sqrt{ (s/2+{\mathbf p}_i \cdot {\mathbf p}_B)^2/E_i^2- {\mathbf p}_B^2}\,\, ,
\end{eqnarray}
where $\sqrt{s}$ is the center of mass energy, $E_{B}^{*}$ is reconstructed $B$ meson energy 
in the center of mass system, and the four-momenta ${\mathbf p}_B$
and $(E_i,\, {\mathbf p}_i)$ of the reconstructed $B$ and the initial state respectively are 
defined in the laboratory frame.
Correctly reconstructed $B$ mesons show a two-dimensional peak around $m_{ES}$ equal
to the $B$ mass and $\Delta{E}=0$. The DIRC information and ${\mathrm d}E/{\mathrm d}x$
are used to identify charged Kaons and pions. The direct CP violation is 
measured as:
\begin{equation}
{\cal A}_{K\pi} = \frac{ n_{K^-\pi^+} - n_{K^+\pi^-} }{ n_{K^-\pi^+} + n_{K^+\pi^-} }\,\, .
\end{equation}
We measure ${\cal A}_{K\pi} = -0.133 \pm 0.030 \stat \pm 0.009 \syst$~\cite{Kpi}, 
which corresponds
to a $4.2\sigma$ deviation from zero. As a comparison, in the decay $B^+ \rightarrow K^+\pi^0$,
which may also exhibit a large asymmetry~\cite{Kpi0CP}
we measure an asymmetry of $0.09\pm 0.09 \stat \pm 0.01 \syst$~\cite{Kpi0}.

\subsection{Time-dependent CP violation and $\sin 2\beta$}

CP violation in the interference between mixing and decay can be observed
as a time-dependent oscillation of the CP asymmetry.
The amplitude of the oscillation in charmonium 
decay modes provides a theoretically
clean determination of the parameter $\sin 2\beta$ of the unitarity triangle. 
Combining the measurements in a number of $(c\bar c)K^0$ final states, 
with both $K_S$ and $K_L$, BaBar has measured~\cite{sin2b_cc}:
\begin{equation}
\sin 2\beta = 0.722 \pm 0.040 \stat \pm 0.023 \syst \,\, .
\end{equation}

Using the cleanest modes with decays containing $K_S$, we can observe 
no evidence of a direct CP violation contribution from the
compatibility with measurement of the parameter $|\lambda | = |\bar A/A |$, where $A$
is the $B$ meson decay amplitude, with the unity:
\begin{equation}
|\lambda | = 0.950 \pm 0.031\stat \pm 0.013 \syst \,\, .
\end{equation}

An independent constraint on $\cos 2\beta$ can be obtained from the
decay $B\rightarrow J/\psi K\pi$ using the interference between the $K\pi$
$S$-wave and $P$-wave amplitudes from $K^*\rightarrow K\pi$ decays. We can constrain
the sign of $\cos 2\beta$ to be positive with a confidence level of 86\%~\cite{cos2b}.

The Standard Model predicts other $B$ decay modes, dominated by a single penguin 
amplitude, such as $B^0\rightarrow \phi K^0_S$
and $B^0\rightarrow K^0_S \pi^0$, to have a 
time-dependent CP asymmetry of magnitude $\sin 2 \beta$. A discrepancy
of the CP asymmetry measurements in such channels from the value of 
$\sin 2\beta$ measured in charmonium modes may be due to the
presence of non standard particles running in the penguin 
diagram loops. BaBar has measured such time-dependent amplitudes
in a number of modes~\cite{sin2b_peng}. The results are reported in Table~\ref{tab:sin2b_peng}.
The measured values tend to be systematically lower than the value 
of $\sin 2\beta$ measured with charmonium modes, their average 
being about 2.7 standard deviations lower. The analysis of the 
data that will be collected in the forthcoming runs is necessary
to confirm or not such discrepancy.

\begin{table}[ht]
\begin{center}
\begin{tabular}{|l|l|}  
\hline
Decay mode & CP amplitude \\
\hline
$\phi K^0$          & $0.50\pm 0.25 ^{+0.07}_{-0.04}$ \\
$\eta^\prime K_S^0$ & $0.27\pm 0.14 \pm 0.030$\\
$f^0 K_S^0$         & $0.95^{+0.23}_{-0.32} \pm 0.10$ \\
$\pi^0 K^0_S$       & $0.35^{+030}_{-0.33} \pm 0.04 $ \\
$K^+K^-K^0_S$       & $0.55\pm 0.22 \pm 0.12 $ \\
\hline
$s$-penguin average & $0.42 \pm 0.10$ \\
$(c\bar c)$ average & $0.726\pm 0.037$ \\
\hline
\end{tabular}
\caption{``$\sin 2\beta$'' measurements in different channels. When two errors are 
quoted they refer to the statistical and systematics contribution respectively.}
\label{tab:sin2b_peng}
\end{center}
\end{table}

\subsection{Measurements of $\alpha$}

The most promising way of measuring $\sin 2\alpha$ is through the time-dependent
CP asymmetry of $b\rightarrow u$ tree-level transitions, such as in $B^ì\rightarrow \pi^+\pi^-$.
Such decay suffer from the pollution of penguin contributions that, unlike the
case of charmonium modes, don't have the same week phase as the tree diagram.
In the case of a negligible penguin contribution the time-dependent CP asymmetry
would have a sinusoidal oscillation with an amplitude equal to $\sin 2\alpha$;
with the introduction of penguin contribution, such amplitude is changed 
into $\sqrt{1-C}\sin 2 \alpha_{eff}$, where $C$ is the amplitude of a cosine 
term which is proportional to $\sin \delta$, where $\delta$ is the relative 
penguin strong phase with respect to tree amplitude. 

We have determined~\cite{sin2alpha_pipi} the sine and cosine term amplitudes
of the time-dependent oscillation of the CP-asymmetry for $B^ì\rightarrow \pi^+\pi^-$
to be:
\begin{eqnarray}
{\cal S}_{\pi^+\pi^-} & = & -0.30\pm  0.17\stat\pm 0.03 \syst \,\, , \\
{\cal C}_{\pi^+\pi^-} & = & -0.09\pm  0.15\stat\pm 0.04 \syst \,\, .
\end{eqnarray}

From the study of the complementary isospin channels $B^\pm\rightarrow\pi^\pm\pi^0$ and
$B^0\rightarrow\pi^0\pi^0$ it is possible to determine
an upper bound to $|\alpha -\alpha_{eff}|$. 
We measure~\cite{pippi0} a branching fraction 
${\cal B}(B^\pm\rightarrow\pi^\pm\pi^0)=(5.80\pm 0.06\stat \pm 0.40\syst)\times 10^{-6}$,
and the direct CP asymmetry for the same channel is 
${\cal A}_{\pi^\pm\pi^0}=-0.01\pm 0.10\stat\pm 0.02\syst$.
For the $\pi^0\pi^0$ channel we measure~\cite{pi0pi0}
${\cal B}(B^0\rightarrow\pi^0\pi^0)=(1.17\pm 0.32\stat \pm 0.10\syst)\times 10^{-6}$,
and a cosine amplitude for the time-dependent CP asymmetry to be:
${\cal C}_{\pi^0\pi^0}=-0.12\pm 0.56\pm 0.06$.
Those measurements allow, using Ref.~\cite{gronau_london} ,
to set the upper limit, at 90\% confidence level:
\begin{equation}
|\alpha -\alpha_{eff}| < 35^\circ\,\, .
\end{equation}

A more favourable situation has been found for 
the measurement of $\alpha$ with $B^0\rightarrow\rho^+\rho^-$.
This channel requires an angular analysis of the final state, 
because it is not a CP eigenstate. From a measurement
of the polarization it turns out that the state is completely
longitudinally polarized: 
$f_{long}=1.00\pm 0.02$, which 
corresponds to a pure CP=1 state. 
From an update using $277\times 10^6$ $B\bar B$ events 
of the analysis from Ref.~\cite{rhoprhom}., 
the time-dependent analysis yields the measurement of 
the sine and cosine amplitudes:
\begin{eqnarray}
{\cal S}_{long} & = & -0.19\pm \stat 0.33\pm 0.11\syst\,\, \label{eq:srhorho}, \\
{\cal C}_{long} & = & -0.23\pm \stat 0.24\pm 0.14\syst\,\, \label{eq:crhorho},
\end{eqnarray}
Where the last limit has been updated with $122\times 10^6$ $B\bar B$ events~\cite{rho0rho0}
with respect to Ref.~\cite{rho0rhop}.
The measurements of the branching fractions of the
corresponding isospin channels are:
\begin{eqnarray}
{\cal B}(B^\pm\rightarrow\rho^\pm\rho^0)&=&(22.5^{+5.7}_{-5.4}\stat \pm 5.8\syst )\times 10^{-6}\,\, ,\label{eq:rhorho0}\\
{\cal B}(B^0\rightarrow\rho^0\rho^0) & < & 1.1\times 10^{-6} {\mathrm (90\% {\mathrm C}.{\mathrm L}.)}\,\,\label{eq:rhorho} .
\end{eqnarray}

Using (\ref{eq:srhorho}, \ref{eq:crhorho}) we can determine $\alpha$ to be:
\begin{equation}
\alpha = (96 \pm 10 \stat \pm 4 \syst \pm 11 ({\mathrm p}{\mathrm e}{\mathrm n}{\mathrm g}.) )^\circ\,\, ,
\end{equation}
where the upper limit to the penguin uncertainty has been 
determined from (\ref{eq:rhorho0}, \ref{eq:rhorho}) using the Grossman-Quinn bound~\cite{grossman_quinn}.

\subsection{Studies for the measurement of $\gamma$}

Different approaches\cite{gronau_etal} have been suggested to determine
the unitarity angle $\gamma$ using the interference a $b\rightarrow c$ 
and a $b\rightarrow u$ transition, whose relative phase is related to $\gamma$, 
in the decays $B^-\rightarrow D^{(*)0}K^-, \bar D^{(*)0}K^-$
with subsequent decays into final states accessible to both charmed meson and anti-meson.
One of the main problem from the experimental point of view
is that the size of the CP asymmetries involved depend on the ratio
of the favoured and the color suppressed decays:
\begin{equation}
r^{(*)}_B = \left|\frac{{\cal A}(B^-\rightarrow \bar D^{(*)0}K^-)}{{\cal A}(B^-\rightarrow D^{(*)0}K^-)}\right|\,\, ,
\end{equation} 
which is expected to be in the range $0.1\div 0.3$.
Using the method suggested by Atwood, Dunietz and Soni, we have studied the
subsequent decays of $D^0$ and $\bar D^0$ in $K^+\pi^-$, where one can assume
as input\cite{babar_rd}:
\begin{equation}
r_D = \left|\frac{{\cal A}(D^0\rightarrow K^+\pi^-)}{{\cal A}(D^0\rightarrow K^-\pi^+)}\right| = 0.060\pm 0.003\,\, .
\end{equation} 
No signal has been observed~\cite{ADS}, allowing to set the limits $r_B<0.23$ and 
$r_B^*<0.16$ at the 90\% confidence level.

Another approach~\cite{gamma_dalitz_th} to the measurement of gamma 
involves the Dalitz analysis of the three-body decay $D^0\rightarrow K_S\pi^-\pi^+$.
This method has the advantage to involve the entire resonant substructure of the three-body 
decay, with the interference of both Cabibbo-allowed and doubly Cabibbo-suppressed amplitudes.
The results of the analysis\cite{gamma_dalitz} may be expressed as the limits:
$r_B<0.24$ and $r_B^*<0.18$ at the 90\% confidence level.
The constraint on $\gamma$ is still rather limited: 
$\gamma =(88\pm 41 \stat \pm 19 \syst \pm 10 ({\mathrm m}{\mathrm o}{\mathrm d}.))^\circ$,
the latter error reflecting the uncertainty in the Dalitz model.

The decays modes $B^0\rightarrow D^{(*)-}\pi^+, D^{(*)-}\rho^+$
receive contributions from a favoured $b\rightarrow c$
and a suppressed $b\rightarrow u$ amplitudes, whose interference
is related to $\sin(2\beta+\gamma)$, that could be 
measured from the time-dependent CP asymmetry.
The limiting experimental factors are the small amount of the asymmetry.
We have studied $B^0\rightarrow D^{(*)-}\pi^+$ with both
full reconstruction
of the final state~\cite{sin2bg_full} and partial
reconstruction, with the $D^{(*)-}$ being tagged with the identification 
of the soft pion only~\cite{sin2bg_part}.

\section{Semileptonic $B$ decays}

\subsection{Measurements of $|V_{cb}|$}

We have measured the exclusive branching fraction of the
decay $\bar B^0\rightarrow D^{*+}\ell^-\bar\nu_\ell$, whose magnitude
is proportional to $|V_{cb}|$. 
The branching fraction has been determined to be, averaging over $\ell=e,\,\mu$~\cite{vcb_dstarlnu}:
\begin{equation}
{\cal B}(\bar B^0\rightarrow D^{*+}\ell^-\bar\nu_\ell) = (4.90 \pm 0.07 \stat ^{+0.36}_{-0.35} \syst)\times 10^{-3}\,\,\ .
\end{equation}
The differential decay rate can be measured as a function of $w$, the relativistic boost 
of the $D^{*+}$ in rest frame of the $B^0$. In the limit of infinite $b$-quark and 
$c$-quark masses, the differential decay rate can be determined from a single Isgur-Wise
function~\cite{isgur_wise}:
\begin{equation}
\frac{{\mathrm d}\Gamma}{{\mathrm d}w} \propto {\cal G}(w){\cal F}(w)^2|V_{cb}|^2\,\, .
\end{equation}
The value of the form factors ${\cal G}(w){\cal F}(w)^2$ at $w=1$ can be predicted 
using lattice calculations~\cite{vcb_lattice}.
The differential rate has been fitted to the experimental data
using a Taylor expansion, and the result of the fit extrapolated to the value $w=1$, 
providing the following measurement of $|V_{cb}|$:
\begin{equation}
|Vcb| = ( 38.7 \pm 0.3 \stat \pm 1.7\syst ^{+1.5}_{-1.3}\theor ) \times 10^{-3}\,\, .
\end{equation}

Another more accurate measurement of $|V_{cb}|$ can be obtained
extracting from the distributions of hadron mass and lepton energy
spectra the moments in inclusive decays $\bar B\rightarrow X_c\ell^-\bar\nu_\ell$, defined as:
\begin{equation}
M_0 = \frac{ \int_{ x_{cut} }^{\infty} {\mathrm d}\Gamma}{ \Gamma_B}\,\, ,
M_1 = \frac{ \int_{ x_{cut} }^{\infty} x {\mathrm d}\Gamma}{ \int_{ x_{cut} }^\infty {\mathrm d}\Gamma}\,\, ,
M_n = \frac{ \int_{ x_{cut} }^{\infty} ( x - M_1 )^n {\mathrm d}\Gamma}{ \int_{ x_{cut} }^\infty {\mathrm d}\Gamma}\,\, , (n=2,3)\,\, , \label{eq:vcb_mom}
\end{equation}
where the variable $x$ is either the hadron mass or the lepton energy. The 
moments defined in (\ref{eq:vcb_mom}) can be related via Operator
Product Expansions (OPE)~\cite{vcb_ope} to fundamental parameters
of the Standard Model including as the CKM matrix elements $|V_{cb}|$ and $|V_{ub}|$
and the heavy quark masses $m_b$ and $m_c$. The fit of the dependency 
of the moments on the applied cut to the hadron mass or the lepton energy~\cite{vcb_mom}
leads to the determination of a number of observables. In particular,
we determine~\cite{vcb_fit}:
\begin{eqnarray}
{\cal B} (b\rightarrow c\ell\nu) & = & (10.61 \pm 
                                          0.16 ({\mathrm e}{\mathrm x}{\mathrm p}.) \pm 
                                          0.06({\mathrm t}{\mathrm h}., {\mathrm H}{\mathrm Q}{\mathrm E}))\%\,\, , \\
|V_{cb}| & = & (41.4\pm 0.4 ({\mathrm e}{\mathrm x}{\mathrm p}.)\pm 0.4({\mathrm H}{\mathrm Q}{\mathrm E})\pm 0.6 \theor )\times 10^{-3}\,\, .
\end{eqnarray}
The errors refer to the experimental, HQE, and additional theoretical uncertainties.
We could have with the same fit precise determination of the $b$ and $c$ quark masses:
\begin{eqnarray}
m_b(1\GeV) & = & (4.61 \pm 0.05 ({\mathrm e}{\mathrm x}{\mathrm p}.) \pm 0.04 ({\mathrm H}{\mathrm Q}{\mathrm E}) \pm 0.02 \theor )\GeV /c^2\,\, ,\\
m_c(1\GeV) & = & (1.18 \pm 0.07 ({\mathrm e}{\mathrm x}{\mathrm p}.) \pm 0.06 ({\mathrm H}{\mathrm Q}{\mathrm E}) \pm± 0.02 \theor )\GeV /c^2\,\, .
\end{eqnarray}
The $c$ quark mass measurement is in very good agreement with the
theoretical prediction using QCD Spectral Sum Riles (QSSR) from 
the measured $D(0^-)$ and $D_s(0^-)$ masses:
$\bar  m_c(m_c)=1.13^{+0.08}_{-0.04}$~GeV$/c^2$~\cite{narison_mc}.

\subsection{Measurements of $|V_{ub}|$}

We have measured $|V_{ub}|$ from the study of inclusive electron 
spectrum in $B\rightarrow X_u e\nu$ decays near the kinematic limit
accessible to $B\rightarrow X_c e\nu$ decays.
The partial branching fraction has been measured in electron momentum range from 2.0 to
2.6 GeV$/c$ and has been extrapolated to the full momentum range
using~\cite{vub_extrap} a previous measurements of the inclusive photon spectrum in $B\rightarrow X_s \gamma$ decays~\cite{vub_bsg}.
The result is~\cite{vub_endpoint}:
\begin{equation}
|V_{ub}| = (3.94 \pm 0.25 (exp) \pm 0.37 \pm (f_u) \pm 0.19 (th) ) \times 10^{-3}\,\, ,
\end{equation}
where the first error is the sum in quadrature of the statistical and systematic uncertainties,
the second refers to the uncertainty of the determination of the fraction $f_u$ of the inclusive 
electron spectrum in the range from 2.0 to 2.6 GeV$/c$, and the third error is due to
theoretical uncertainties in the QCD corrections and the $b$-quark mass.

Further measurements of $|V_{ub}|$ can be obtained
from the studies of exclusive $b\rightarrow u\ell\nu$ 
decay channels. BaBar has developed a novel technique
to approach such decays, by completely reconstructing
the decay of the companion $B$ meson, thus significantly
reducing the background to the reconstruction of the
semileptonic decay. The price to pay is the reduction
of statistics, that will anyway be overcome with the increase
of collected data sample, when the statistical sensitivity
will approach the systematic limit.
Preliminary measurements of exclusive branching fractions are~\cite{vub_excl}:
\begin{eqnarray}
{\cal B}(B^0\rightarrow \pi^-\ell^+\nu) & = & (1.08\pm 0.28 \stat \pm 0.16 \syst)\times 10^{-4}\,\, ,\\
{\cal B}(B^+\rightarrow \pi^0\ell^+\nu) & = & (0.91\pm 0.28 \stat \pm 0.14\syst)\times 10^{-4}\,\, ,\\
{\cal B}(B^0\rightarrow \rho^-\ell^+\nu) & = & (2.57\pm 0.52 \stat \pm 0.59\syst)\times 10^{-4}\,\, .
\end{eqnarray}
The extraction of $|V_{ub}|$ is in progress.

\section{Channels probing new physics}

$B$ decays provide a probe to explore new physics processes which could arise 
with the exchange of virtual non-Standard particles.

\subsection{Radiative penguin decays}

Radiative $b\rightarrow s\gamma$ and $b\rightarrow d\gamma$ decays proceed 
in the Standard Model via electromagnetic penguin diagrams. New particles
could replace the $W$ and quarks exchanges in the penguin loop 
producing deviation in the rate and CP asymmetries with respect to
the Standard Model predictions.
While the world average of the decay rate of $b\rightarrow s\gamma$ 
($(3.3\pm 0.4)\times 10^{-4}$~\cite{bsgamma_pdg}), 
in agreement with the Standard Model prediction~\cite{bsgamma_sm},
has reached a level of uncertainty comparable to the
error on the theoretical prediction, the $b\rightarrow d\gamma$
are at the limit of discovery according to the Standard Model
prediction. BaBar has set the following 90\% confidence level limits~\cite{bdgamma_limits}:
\begin{eqnarray}
{\cal B}(B\rightarrow\rho^+ \gamma) & < & 1.8 \times 10^{-6}\,\, ,\\
{\cal B}(B\rightarrow\rho^0 \gamma) & < & 0.4 \times 10^{-6}\,\, ,\\
{\cal B}(B\rightarrow\omega\gamma)  & < & 1.0 \times 10^{-6}\,\, ,
\end{eqnarray}
that can be combined into:
\begin{equation}
{\cal B}(B\rightarrow (\rho/\omega) \gamma)<1.2 \times 10^{-6}\,\, .
\end{equation}

The study of CP asymmetry of $b\rightarrow s\gamma$ has been performed on 
a number of inclusive final states containing one charged or neutral kaon and
one to three pions~\cite{bsgamma_cp}. The measured asymmetry leads to:
\begin{equation}
{\cal A}_{CP}(b\rightarrow s\gamma) = 0.025\pm 0.050 \stat \pm 0.015 \syst\,\, ,
\end{equation}
to be compared to a Standard Model prediction of less than 1\% 
~\cite{bsgamma_cp_sm}.
In the exclusive channel $B\rightarrow K^*\gamma$ the CP asymmetry
is constrained in the range:
\begin{equation}
-0.074 < {\cal A}_{CP}(B\rightarrow K^*\gamma) < 0.049\,\, ,
\end{equation}
at 90\% confidence level.

\subsection{Search for pentaquarks}

Recent experimental studies have been reported observations of new exotic
baryon resonances with narrow width which could be interpreted as states
composite of five quarks. In particular the $\Theta^+$ with a mass around 1540~MeV$c^2$
has been reported in Ref.~\cite{penta_theta1540}, the $\Xi^{--}$ 
and $\Xi^{0}$, both with masses around 1862~MeV$c^2$  have been reported in 
Ref.~\cite{penta_csimm} and the $\Theta^0_c$ with a mass of 3099~MeV$c^2$
has been reported in Ref.~\cite{penta_theta0c}. Several theoretical
models have been proposed to explain such states~\cite{penta_th}.

Pentaquark states may be produced in $e^+e^-$ collisions as well, and
the very large statistics collected by BaBar may provide a
way to probe pentaquark production down to very low branching fractions.
We have searched for the $\Theta^{*++}$ in the decay $B^+ \rightarrow \Theta^{*++} \bar p$
where $\Theta^{*++}\rightarrow p K^+$ using 81~fb$^{-1}$ of data~\cite{penta_theta_starpp} and
we set an upper limit on the branching fraction to be $1.5\div 3.3\times10^{-7}$,
at 90\% confidence level, depending on the mass of the $\Theta^{*++}$, 
which has been assumed to vary from 1.43 to 2.00 GeV$/c^2$.
We have also performed an inclusive search for strange pentaquark production 
using 123~fb$^{-1}$ of data. Different decay channels have been studied,
assuming the quark content of the $\Theta^+(1540)$ to be $udud\bar s$, and
of the $\Xi^{--}(1860)$ and $\Xi^0(1860)$ to be $dsds\bar u$ and $uss(u\bar u+d\bar d)$
respectively. In addition we have searched for other members of the antidecuplet
and corresponding octet that would complete the five-quark model 
for such states. Though we found very clear signal for known barions, 
demonstrating the experimental sensitivity to narrow resonances in the mass
range of interest, we found no evidence for the production of
pentaquark states. We set a number of limits on their
production cross sections as functions their of center of mass momentum. 
The complete list of results can be found in Ref.~\cite{penta_theta_p1540};
the limits are at the level of $10^{-4}\div 10^{-5}$ per event, depending on the width assumed,
valid for any narrow state with a mass close to the range $1540\div 1860$~MeV$c^2$.

\section{Study of charmed meson spectroscopy}
In 2003 BaBar discovered the $D_{sJ}^*(2317)^+$ meson~\cite{dsj_2317}, 
confirmed by other observations~\cite{dsj_cleo,dsj_confirm}.
Subsequently the $D_{sJ}(2460)^+$ meson was also observed~\cite{dsj_cleo,dsj_babar}.
The two discoveries reawakened interest in charmed meson spectroscopy.
The spectroscopy of $c\bar s$ states can be described 
in the limit of large charm-quark mass~\cite{csbar_spectr}. 
Under that limit, the sum of the orbital and spin momenta 
$\overrightarrow{j} =\overrightarrow{l}+\overrightarrow{s}$ 
is conserved. The positive-parity $P$-wave states have $j = 3/2$ or 
$j = 1/2$. Combining those two states with the spin 
of the heavy quark, we have, from the state with $j = 3/2$,
the possible values of the total angular 
momentum $J = 2$ and $J = 1$, and from the state with $j = 1/2$
the values $J = 1$ and $J = 0$. 
The members of the $j = 3/2$ doublet 
are expected to have small width~\cite{csbar_mass_1}, 
the state $J^P = 2^+$ being identified to be the $D^*_{sJ} (2573)^+$,
while the state $J^P = 1^+$ is identified with the $D_{s1}(2536)^+$.
The observed narrow width of the $D_{sJ}^*(2317)^+$ and $D_{sJ}(2460)^+$,
which are below the kinematical threshold of the decay by kaon emission,
are in contradictions with some prediction
of states with masses between 2.4 and 2.6~GeV$/c^2$~\cite{csbar_mass_1,csbar_mass_2}, which would also 
have a large widths because of the dominant decays to $D^{(*)}K$.
A review of recent the theoretical approaches towards the
computation of the masses of the new states, including the hypothesis of
unconventional multiquark states can be found in Ref.~\cite{ds_colangelo}.
Different models also provide prediction for decay branching ratios.
There is hence experimental interest in the determination of the properties of the 
newly discovered states. Detailed studies of their decays can now provide 
more accurate information.
Using the decay $D_{sJ}^*(2317)^+\rightarrow D_s^+\pi^0$ we obtain the mass measurement~\cite{dsj_mass}:
\begin{equation}
m(D_{sJ}^*(2317)^+) = 2318.9\pm 0.3 \stat \pm 0.9 \syst \,\, \MeV/c^2\,\,.
\end{equation}
Averaging the measurements from the decays of the $D_{sJ}(2460)^+$ to $D_s^+\gamma$, $D_s^+\pi^0\gamma$ 
and $D_s^+\pi^+\pi^-$ we obtain:
\begin{equation}
m(D_{sJ}(2460)^+) = 2459.4\pm 0.4 \stat \pm 1.2 \syst \,\, \MeV/c^2\,\,.
\end{equation}
The relative branching fractions of the decays under study are also measured
with an uncertainty of the order of $15\div 20$\%.

The study of the distribution of helicity angle of the decay $D_{sJ}(2460)^+\rightarrow D_s^+\gamma$
in decays $B\rightarrow D_{sJ}(2460)^+\bar D^{(*)}$ can be used to obtain information
on the $D_{sJ}(2460)^+$ spin $J$~\cite{dsj_hel}. BaBar observations
favours the hypothesis of a $J^P=1^+$ state, excluding a $J^P=2^+$ state.
A recent calculation~\cite{ds_narison}, using QCD spectral sum rules,
which are less affected by large $1/m_c$ corrections than HQET, also
including radiative corrections, provides the estimate 
$m(D^*_s(1^+)) = (2440 \pm 113)$~MeV$/c^2$, in agreement with our measurement
of the $D_{sJ}(2460)^+$ mass, supporting a $J^P=1^+$ assignment.

\bigskip
\section{Acknowledgements}
We are grateful for the extraordinary contributions of our 
PEP-II colleagues in achieving the
excellent luminosity and machine conditions that have made 
this work possible. The success of
this project also relies critically on the expertise and 
dedication of the computing organizations
that support BABAR. The collaborating institutions wish to 
thank SLAC for its support and the
kind hospitality extended to them. This work is supported by 
the US Department of Energy and
National Science Foundation, the Natural Sciences and Engineering 
Research Council (Canada),
Institute of High Energy Physics (China), the Commissariat {\`a} 
l' Energie Atomique and Institut
National de Physique Nucl{\'e}aire et de Physique des 
Particules (France), the Bundesministerium f\"ur
Bildung und Forschung and Deutsche Forschungsgemeinschaft (Germany), the Istituto Nazionale
di Fisica Nucleare (Italy), the Foundation for Fundamental Research on Matter (The Netherlands),
the Research Council of Norway, the Ministry of Science and Technology of the Russian Federation,
and the Particle Physics and Astronomy Research Council (United Kingdom). Individuals have
received support from CONACyT (Mexico), the A. P. Sloan Foundation, the Research Corporation,
and the Alexander von Humboldt Foundation.

\end{document}

%% file: econfmacros.tex
\def\beq{\begin{equation}}
\def\eeq#1{\label{#1}\end{equation}}
\def\eeqn{\end{equation}}

\def\beqa{\begin{eqnarray}}
\def\eeqa#1{\label{#1}\end{eqnarray}}
\def\eeqan{\end{eqnarray}}

\let\bar=\overbar

\def\etal{{\it et al.}}

\def\Dslash{\not{\hbox{\kern-4pt $D$}}}
\def\dslash{\not{\hbox{\kern-2pt $\del$}}}

\def\msb{{\bar{\ssstyle M \kern -1pt S}}}

